\documentclass[preprints,article,accept,moreauthors,pdftex]{Definitions/mdpi} 

\firstpage{1} 
\pubvolume{xx}
\issuenum{1}
\articlenumber{5}
\pubyear{2019}
\copyrightyear{2019}
\history{Received: date; Accepted: date; Published: date}

\usepackage{upgreek}

\Title{Proton CT -- a novel diagnostic tool in cancer therapy}

\Author{Monika Varga-Kofarago $^{1}$\orcidA{} on behalf of the Bergen pCT collaboration}

\AuthorNames{Firstname Lastname, Firstname Lastname and Firstname Lastname}

\address{%
$^{1}$ \quad HAS Wigner Research Centre for Physics; varga-kofarago.monika@wigner.mta.hu}

\abstract{Radiotherapy is one of the main methods in the successful treatment of cancer. The tumor is irradiated with photons or charged particles (e.g.\ protons), and in the case of massive charged particles, the treatment results in less unnecessary dose outside the tumor and therefore less side effects for the patient and a faster recovery. However, the dose planning of hadron therapy is calculated from photon CT measurements, which results in large uncertainties in the planning and therefore in a necessary enlargement of the treatment area. This uncertainty can be reduced by performing the CT scan using protons. The current contribution shows the development of a sampling calorimeter for proton CT measurements and describes the state of the project.}


\begin{document}

\section{Introduction}
\begin{figure}[H]
\centering
\includegraphics[width=0.5\textwidth]{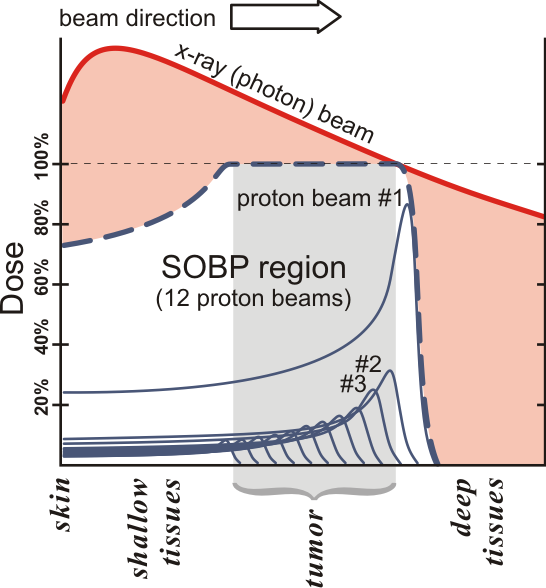}
\caption{Relative dose of photons and protons as a function of the depth in the tissue~\cite{bragg}.}
\label{fig:relative_dose}
\end{figure}

As shown in Ref.~\cite{stat}, around 40\% of the population is effected by cancer at a certain point in their life. Radiotherapy is one of the main successful methods in treating cancer~\cite{radiotherapy}. In radiotherapy, photons or hadrons can be used to irradiate the patient. The goal of these treatments is to damage the DNA of the cancer cells, while causing as little damage as possible to the healthy tissues. This damage can be minimized by irradiating the patient from several angles, therefore delocalizing it. If it is possible to focus the radiation on the tumor, the damage can be further minimized. If photons are used, only the delocalization can be applied as photons are absorbed mostly at their entrance to the patient (see Figure~\ref{fig:relative_dose}). However, if protons or other charged particles are used, the damage can be focused inside the tumor by adjusting the energy of the beam and positioning the Bragg peak in the tumor. 

\section{Computer tomography with protons}
The problem of current hadron therapy treatments is that the treatment is planned after acquiring an image of the patient by a photon CT scan. This results in large uncertainties (around 3--4\%) in the determination of the stopping power of protons in front of the tumor~\cite{error}. This is due to the fact that the relation of the attenuation coefficients of photons and the stopping power of protons is not linear and not one-to-one as it differs depending on the type of the tissue~\cite{conversion}. 

\begin{figure}[H]
\centering
\includegraphics[width=0.5\textwidth]{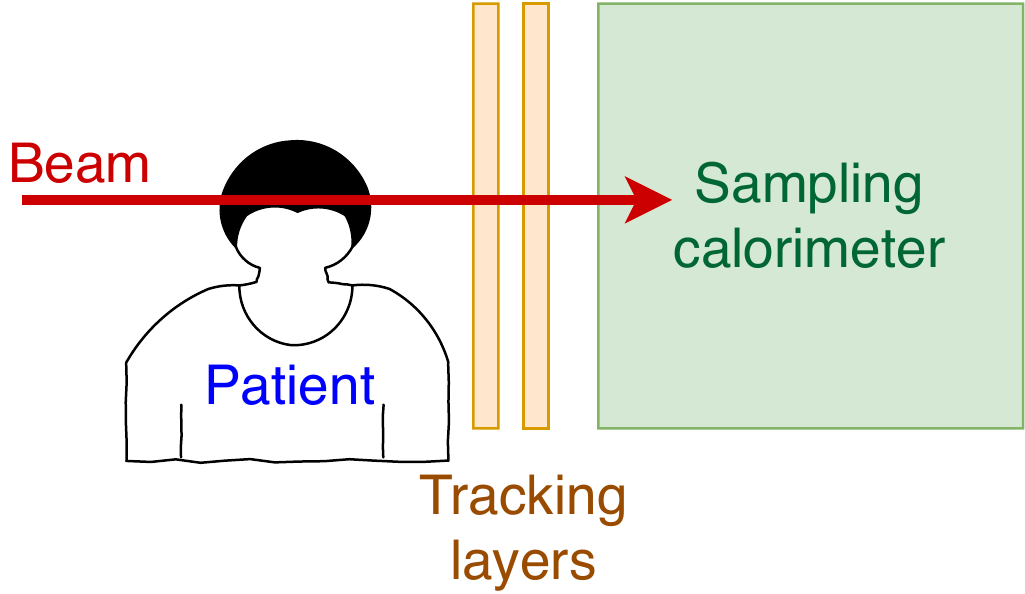}
\caption{The concept of a proton CT detector.}
\label{fig:concept}
\end{figure}

This problem can be solved by using protons for the imaging in the CT measurement instead of photons, therefore the measurement will give directly the stopping power for protons. This would reduce the uncertainty by more than a magnitude to 0.3\%~\cite{error}. Such a measurement would use protons with a higher energy than the ones used for the treatment, such that their Bragg peak would fall outside of the patient and in the detector placed behind the patient. The position of the protons has to be determined before entering the patient and after leaving the patient, and after the patient the energy of the protons has to be measured as well. Before the patient, the position of the protons can be determined by the measurement of the beam position or by a tracking detector with very low material budget (maximum 50--100~$\upmu$m of silicon). After the patient the position and the energy measurement can be achieved by a high resolution sampling calorimeter. The concept of such a detector can be seen in Figure~\ref{fig:concept}. If the measurement is done prior to the treatment, it can be used for the planning of the treatment, while if it is done quasi-simultaneously, it can be used for dose verification, dose optimization or patient alignment. Patient alignment is not a trivial task, as the unavoidable movement of the patient has to be taken into account during the treatment. This can be done by monitoring the patient with a quasi-simultaneous proton CT measurement.

\section{The proposed calorimeter}
\begin{figure}[H]
\centering
\includegraphics[width=0.65\textwidth]{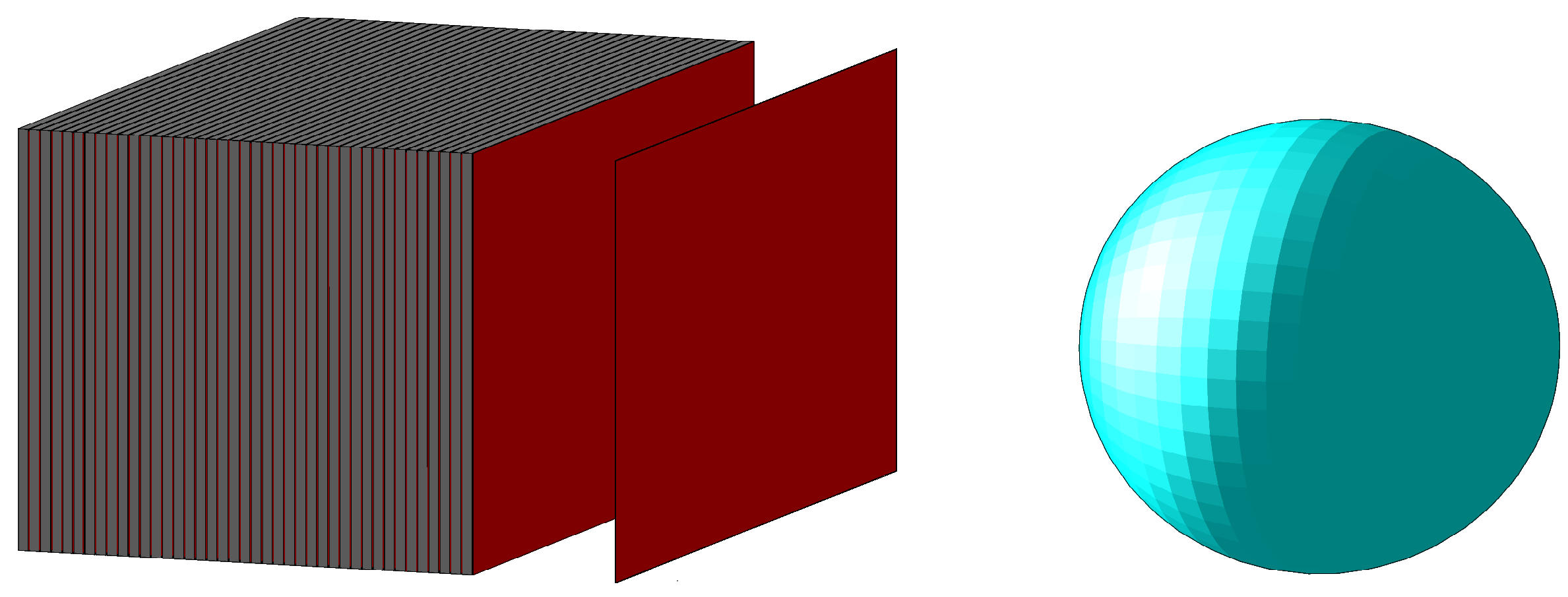}
\caption{The proposed design of the sampling calorimeter. The red layers are the sensitive layers which alternate with the gray aluminum layers. The blue sphere represents the patient.}
\label{fig:optimization}
\end{figure}

The active part of the sampling calorimeter will be the ALPIDE sensor~\cite{ALPIDE} which was originally developed for the Inner Tracking System of the ALICE detector~\cite{TDR}. These ALPIDE layers will alternate with aluminum layers which act as energy degraders for the protons. There will be 41 sensitive layers and 41 degrader layers and each aluminum layer will be 3.5~mm thick. The full front area of the detector will be 27~cm~$\times$~15~cm, which is made up of 9~$\times$~9 ALPIDE sensors. The proposed calorimeter design can be seen in Figure~\ref{fig:optimization}.

The ALPIDE is a digital MAPS type silicon detector. It was designed to function in the ALICE detector, therefore it is radiation tolerant up to $ 1.7\times 10^{13}$~1~MeV~$\text{n}_{\text{eq}}\text{/cm}^2$ non-ionizing dose and up to 2700 krad ionizing radiation. This is important as in the case of the proton CT, the detector will be placed directly in the beam. The ALPIDE has a high detection efficiency ($>99\%$), good spatial resolution ($\sim 5~\upmu$m) and low noise ($<10^{-6}$ hits/event/pixel)~\cite{ALPIDE}. It can be produced in a $50~\upmu$m and a $100~\upmu$m thick version, therefore, if needed, it can be used in the tracker in front of the patient as well.

\section{Results from the prototype}
\begin{figure}[H]
\centering
\includegraphics[width=1\textwidth]{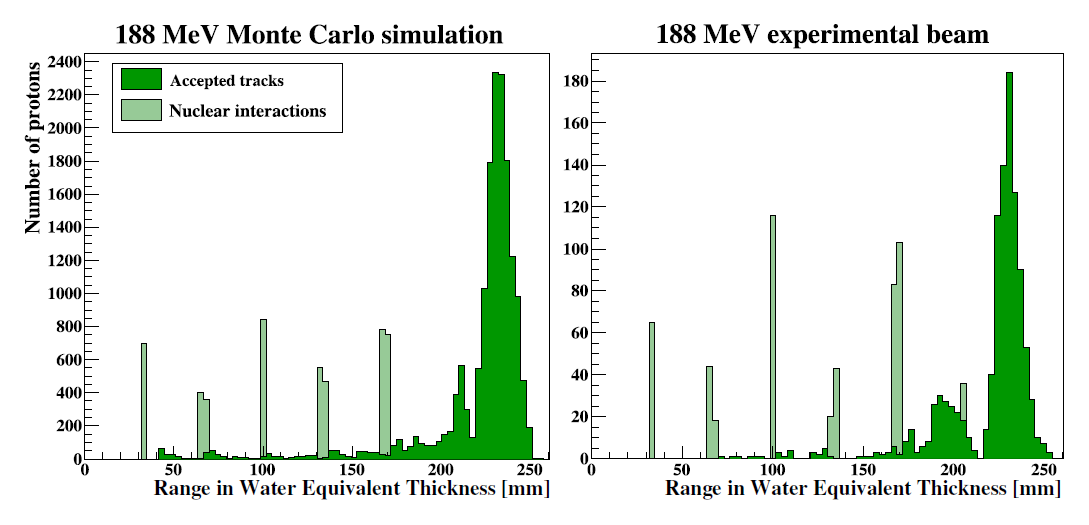}
\caption{Comparison of the simulations (left panel) and the measured results (right panel) of the prototype~\cite{results}.}
\label{fig:mes1}
\end{figure}

\begin{figure}[H]
\centering
\includegraphics[width=0.7\textwidth]{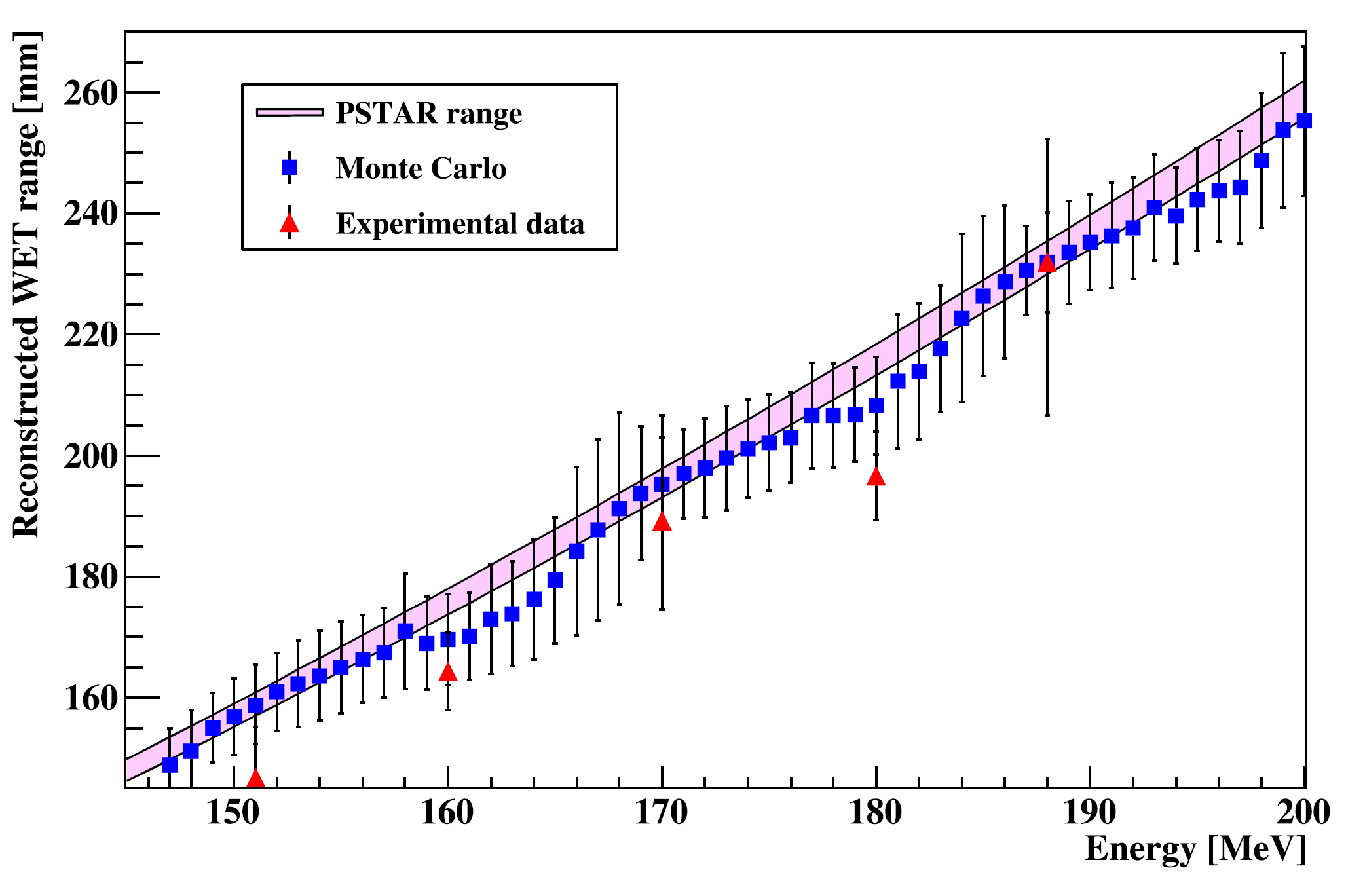}
\caption{Reconstructed range of the protons as a function of the energy of the beam~\cite{results}.}
\label{fig:mes2}
\end{figure}

The first prototype of the calorimeter was not optimized for detecting protons, but for measuring electromagnetic showers. This prototype used MIMOSA23 sensors~\cite{MIMOSA} and used 3.3~mm tungsten absorbers instead of aluminum as a degrader~\cite{Focal}. It was tested in a proton beam at KVI -- Center for Advanced Radiation Technology in Groningen~\cite{KVI}. The comparison of the results with simulations can be seen in Figure~\ref{fig:mes1} which shows the number of reconstructed protons as a function of their reconstructed range. In both cases the Bragg peak is clearly visible around 230~mm, and the simulation (left panel) describes the test beam data (right panel) well. In Figure~\ref{fig:mes2}, the reconstructed range of the tracks is shown as a function of the energy of the beam. The agreement between data and simulation is good here as well, and the observed linear trend shows that the range is a good measurement of the energy of the incoming protons.

\section{Conclusions}
In conclusion, with the help a proton CT the dose estimation of hadron therapy will become more accurate, therefore it will have less side effects and can be applied closer to critical organs. As a proton CT detector a sampling calorimeter of alternating ALPIDE and aluminum layers is proposed. The first prototype of such a detector, which was optimized for electromagnetic showers instead of the energy measurement of protons, shows a good performance, and its performance can be well described by Monte Carlo simulations.

\vspace{6pt}


\funding{This work has been supported by the Hungarian NKFIH/OTKA K 120660 grant.}


\conflictsofinterest{The authors declare no conflict of interest. The funders had no role in the design of the study; in the collection, analyses, or interpretation of data; in the writing of the manuscript, or in the decision to publish the results.} 

\abbreviations{The following abbreviations are used in this manuscript:\\

\noindent 
\begin{tabular}{@{}ll}
  ALICE & A Large Ion Collider Experiment \\
  ALPIDE & ALICE PIxel DEtector \\
  CT & Computer Tomography \\
  DNA & DeoxyriboNucleic Acid \\
  MAPS & Monolithic Active Pixel Sensor \\
  pCT & proton Computer Tomography
\end{tabular}}




\reftitle{References}

\end{document}